\documentclass[11pt]{article}
\usepackage[margin=1in]{geometry}
\usepackage{amsmath,amssymb,amsthm}
\usepackage{mathrsfs}
\usepackage{tabularx}
\usepackage{hyperref}

\newcommand{\scriptI}{\mathscr{I}}

\title{The Principle of Minimum Justified Correlation\\
\large Part I --- Entropy, Fisher Information, and Quantum Kinetic Energy}
\author{John H. Van Drie\\
Van Drie Research, North Andover, Massachusetts\\
\textsf{john@vandrieresearch.com}}
\date{Draft Version $\alpha_2$ --- for discussion\\
Generated Tue., Sept.\ 29, 2026, 12:52 PM ET}

\begin{document}
\maketitle

\section*{Abstract}

It is shown that Shannon entropy, Fisher information, and quantum-mechanical kinetic energy may all be viewed as measures of correlation. We begin with a fundamental correlation-destroying map: a joint probability density $\rho(x,y)$ is replaced by $\rho_x(x)\rho_y(y)$, the product of its marginal distributions. In the discrete case, Shannon's entropy is nondecreasing under this map. For continuous distributions, this fundamental map leads to a well-behaved, coordinate-invariant correlation measure, analogous to the discrete Shannon entropy:
\[
\scriptI[\rho] = h[\rho_x] + h[\rho_y] - h[\rho].
\]
In the continuous case, however, another measure appears: Fisher information. The relative Fisher information $J(\rho\|\rho_x\rho_y)$ behaves similarly under the fundamental map. For a normalized real quantum wavefunction, this decrease is exactly proportional to the decrease in mean kinetic energy,
\[
\langle T\rangle_\psi - \langle T\rangle_\Phi = \frac{\hbar^2}{8m} J(\rho\|\rho_x\rho_y), \qquad \Phi = \sqrt{\rho_x \rho_y}.
\]
In Jaynes's language, the result supports a principle of minimum justified correlation: given physical constraints and a set of possible distributions or related amplitudes satisfying those constraints, select from that set those with the least correlation.

This is Part I of a two-part paper. Here we develop the entropy, Fisher-information, and kinetic-energy identities above, and state the principle they support. Part II addresses questions which Part I raises but leaves unanswered: multiple solutions, time dependence, the role of spin, a route to the Schr\"odinger equation itself, and a proposed experimental test. AI has been used.

\section{Introduction: the fundamental correlation-destroying map for a discrete distribution}
\label{sec:intro}

Consider, as we did earlier \cite{vandrie2000a,vandrie2000b}, two discrete random variables $X$ and $Y$, with joint probability distribution $p_{ij}$. Define the fundamental map:
\[
\boxed{\ C : p_{ij} \longmapsto \pi_{ij} = P_i Q_j, \qquad P_i = \sum_j p_{ij}, \quad Q_j = \sum_i p_{ij}. \ } \tag{1.1}
\]
This map removes the dependence between $X$ and $Y$. Under it, the Shannon entropy can only increase, or stay the same, while leaving both marginal distributions unchanged:
\[
H(p) \le H(\pi) = H(P) + H(Q). \tag{1.2}
\]

Equation (1.2) is simple, but fundamental. Destroying correlation increases entropy, or leaves it unchanged, and the increase is exactly the correlation removed:
\[
H(P) + H(Q) - H(p) = H(\pi) - H(p) = I(X;Y) \ge 0. \tag{1.3}
\]

$I(X;Y)$ is the mutual information in $p$. We use the phrase \emph{decorrelation property} for any functional whose change under the map $C$ obeys a one-sided inequality like this, with equality exactly when the distribution was already a product to begin with -- the direction of the inequality can go either way depending on the functional. Equation (1.2) restates an observation of Jaynes's: removing correlation makes entropy additive, and it can't make it smaller, with equality only when the original distribution is already a product \cite{jaynes2}.

No new equations have been derived yet. But, as a great man once told me, quoting his own work \cite{feynman1949}: ``there is a pleasure in recognizing old things from a new viewpoint.'' The spirit of what follows is to show that a simple, fundamental map -- the correlation-destroying map $C$ -- underlies these separately known results, in the hope that seeing them as one thing rather than three will eventually lead to something new.

\section{The fundamental map applied to continuous distributions}
\label{sec:continuous-map}

We now move from discrete distributions to continuous ones. Let $\rho(x,y)$ be a normalized continuous distribution,
\[
\int\!\!\int \rho(x,y)\,dx\,dy = 1, \tag{2.1}
\]
with marginal distributions
\[
\rho_x(x) = \int \rho(x,y)\,dy, \qquad \rho_y(y) = \int \rho(x,y)\,dx. \tag{2.2}
\]
The same map $C$ carries over unchanged:
\[
\boxed{\ C : \rho(x,y) \longmapsto \rho_x(x)\rho_y(y). \ } \tag{2.3}
\]
The resulting density is still normalized, and still has the same marginals. This is the single transformation used throughout the rest of the paper; everything that follows amounts to asking what this same map does to one quantity after another.

The obvious continuous analog of Shannon entropy is
\[
h[\rho] = -\int\!\!\int \rho(x,y) \log \rho(x,y)\,dx\,dy, \tag{2.4}
\]
usually called differential entropy. It has the decorrelation property we want, but it has a disqualifying defect: unlike discrete Shannon entropy, differential entropy depends on the choice of units. Rescale coordinates, $x'=ax$, $y'=by$ with $a,b\ne0$, and
\[
h[\rho'] = h[\rho] + \log|ab|.
\]
Choosing $|ab|$ small enough makes $h[\rho']$ negative. More generally, differential entropy is not invariant under smooth coordinate changes.

\section{Continuous mutual information emerges from the fundamental map}
\label{sec:mutual-info}

Take the continuous analogs of the four quantities in Eq.~(1.3):
\[
\scriptI[\rho] \equiv h[\rho_x] + h[\rho_y] - h[\rho] = h[C\rho] - h[\rho] = \int\!\!\int \rho \log\frac{\rho}{\rho_x \rho_y}\,dx\,dy \ge 0. \tag{3.1}
\]
Call $\scriptI[\rho]$ the entropy correlation, or continuous mutual information, of $\rho$. Unlike the individual differential entropies that went into it, every expression in Eq.~(3.1) survives separate smooth, invertible changes of the $x$ and $y$ coordinates -- homothetic transformations, translations $x'=x+c$, $y'=y+d$, and, when $x$ and $y$ are vectors, independent rotations $x'=R_xx$, $y'=R_yy$. It does not survive a rotation that mixes $x$ and $y$ together; there is no reason to expect the mutual information to remain fixed under it.

Equation (3.1) is thus the coordinate-invariant counterpart of Eq.~(1.3), vanishing exactly when the variables are independent.

N.B. In the discrete case, Shannon proved $H(p)$ is the \emph{only} measure satisfying some natural requirements -- continuity, symmetry, grouping \cite{shannon}. That uniqueness does not carry over: $\scriptI[\rho]$ is \emph{not} the only such measure once we are in the continuous realm, because continuous distributions can be differentiated. This opens the door to a second correlation measure, translational Fisher information, which shares the decorrelation property and additionally captures how sharply the dependence varies in space.

\section{The Fisher-information change under the fundamental map}
\label{sec:fisher}

The general strategy is the same throughout: take a functional, compare its value before and after applying $C$, and ask whether the difference has a definite sign. We did this for entropy above; we now do it for Fisher information. Recall the translational Fisher information of $\rho$ \cite{fisher1925},
\[
J[\rho] = \int\!\!\int \rho\, |\nabla \log \rho|^2\,dx\,dy = \int\!\!\int \frac{|\nabla \rho|^2}{\rho}\,dx\,dy. \tag{4.1}
\]
and, for the marginals,
\[
J[\rho_x] = \int \frac{|\rho_x'|^2}{\rho_x}\,dx, \qquad J[\rho_y] = \int \frac{|\rho_y'|^2}{\rho_y}\,dy. \tag{4.2}
\]
Define the Fisher \emph{correlation} as the excess over the independent product,
\[
J_{\mathrm{corr}}[\rho] \equiv J[\rho] - J[\rho_x] - J[\rho_y]. \tag{4.3}
\]
This is the Fisher-information counterpart of $\scriptI[\rho]$, and it obeys its own clean identity,
\[
J_{\mathrm{corr}}[\rho] = J(\rho\|\rho_x\rho_y) = \int\!\!\int \rho \left| \nabla \log \frac{\rho}{\rho_x\rho_y} \right|^2 dx\,dy \ge 0. \tag{4.4}
\]
We now have two measures of correlation, worth writing side by side:
\begin{equation}
\begin{aligned}
\text{amount of correlation:} &\quad \scriptI[\rho], \\
\text{spatial sharpness of correlation:} &\quad J_{\mathrm{corr}}[\rho] = J(\rho\|\rho_x\rho_y).
\end{aligned}
\tag{4.5}
\end{equation}
Both measure the same underlying departure from independence, in different units. $\scriptI[\rho]$ is dimensionless and tells us \emph{how much} correlation there is, in total. $J_{\mathrm{corr}}[\rho]$ has dimensions of inverse length squared and tells us \emph{how sharply} that correlation varies in space. There is a cost for that extra spatial information: $J_{\mathrm{corr}}[\rho]$, unlike $\scriptI[\rho]$, is not invariant under arbitrary separate homothetic transformations of $x$ and $y$ -- because it is built from a gradient, it picks up the local scale. It is only invariant under translations and separate rotations. This is not a defect; it is the price of detecting spatial structure rather than only a total. It does mean, however, that $\scriptI[\rho]$ and $J_{\mathrm{corr}}[\rho]$ live under different, only partly overlapping, symmetry groups. (Appendix~\ref{app:fisher-decomp} gives the short proof of the Fisher decomposition.)

\section{The quantum-mechanical kinetic-energy change under the fundamental map}
\label{sec:kinetic}

Take a normalized wavefunction $\psi(x,y)$ with Born density
\[
\rho(x,y) = |\psi(x,y)|^2, \qquad \int\!\!\int |\psi|^2\,dx\,dy = 1, \tag{5.1}
\]
and define the decorrelated reference
\[
\Phi(x,y) = \sqrt{\rho_x(x)\rho_y(y)}. \tag{5.2}
\]
Consider, for concreteness, two equal-mass particles each confined to one dimension. The quantum-mechanical kinetic energy has expectation value
\[
\langle T \rangle_\psi = -\frac{\hbar^2}{2m} \int\!\!\int \psi^* \nabla^2 \psi\,dx\,dy. \tag{5.3}
\]
A point worth noting here, relevant again in Section~\ref{sec:jaynes} and Appendix~\ref{app:lowdin}: the $x/y$ split used throughout has to be tied to something physically real, such as two distinguishable particles, and not to an arbitrary choice of axes for a single particle. It is not a coordinate-free notion. Integrating (5.3) by parts,
\[
\langle T \rangle_\psi = \frac{\hbar^2}{2m} \int\!\!\int |\nabla \psi|^2\,dx\,dy. \tag{5.4}
\]
That is the Dirichlet integral of the wavefunction; we assume the boundary term vanishes, so it can equally well be written $-\int\!\!\int \psi^* \Delta \psi\,dx\,dy$, as in (5.3).

The key identity follows. For a \emph{real} wavefunction, this same Dirichlet integral is one quarter of the Fisher information of the Born density:
\[
J[\rho] = 4\int\!\!\int |\nabla \psi|^2\,dx\,dy. \tag{5.5}
\]
This still holds even when $\psi$ changes sign, since $|\nabla|\psi||=|\nabla\psi|$ almost everywhere. One caveat: $\Phi$ from Eq.~(5.2) is not literally ``$\psi$ with the correlation switched off'' whenever $\psi$ itself has nodes -- it is the unique product amplitude with the right marginal densities, and that is the sense in which the comparison is meant. With that understood,
\[
\langle T \rangle_\psi = \frac{\hbar^2}{8m} J[\rho], \qquad \langle T \rangle_\Phi = \frac{\hbar^2}{8m}\big(J[\rho_x] + J[\rho_y]\big). \tag{5.6}
\]
Subtracting these and using the definition of $J_{\mathrm{corr}}[\rho]$ from Eq.~(4.3) gives the kinetic-energy identity that the map $C$ produces in the quantum-mechanical setting:
\[
\boxed{\ \langle T \rangle_\psi - \langle T \rangle_\Phi = \frac{\hbar^2}{8m} J_{\mathrm{corr}}[\rho] = \frac{\hbar^2}{8m} J(\rho\|\rho_x\rho_y) \ge 0. \ } \tag{5.7}
\]

This quantity, $\hbar^2/8m\cdot J[\rho]$, is familiar under another name. In orbital-free density-functional theory it is called the von Weizs\"acker kinetic-energy functional (written here for a general $n$-dimensional density, of which our two-variable $\rho(x,y)$ is the case $n=2$; $d\tau$ is the $n$-dimensional volume element),
\[
T_W[\rho] = \frac{\hbar^2}{8m} \int \frac{|\nabla \rho|^2}{\rho}\,d\tau = \frac{\hbar^2}{8m} J[\rho],
\]
originally proposed as an approximation to the kinetic energy of an interacting many-electron system, using only the density \cite{sears,hamilton}. For a single-particle state with a real wavefunction, however, $T_W[\rho]$ is not an approximation: it is exactly $\langle T\rangle_\psi$, as shown above. Equation (5.7) can therefore be re-read as a superadditivity statement about the von Weizs\"acker functional under marginal factorization,
\[
T_W[\rho] \ge T_W[\rho_x] + T_W[\rho_y],
\]
with equality precisely when $\rho=\rho_x\rho_y$. In plain language: the map $C$ lowers the von Weizs\"acker kinetic energy by exactly the relative Fisher information of $\rho$ against its own marginals. This connects the present construction to a wider practice of using Fisher information as a kinetic-energy surrogate in density-functional theory, except that here Eq.~(5.7) is exact, not a working approximation, as it is usually taken to be in density-functional theory.

\subsection*{The complex case}

The discussion so far has ignored the phase of a general complex wavefunction $\psi(x,y)$. We now include it. Write $\psi=\sqrt\rho\,e^{i\theta}$, and choose as reference the phase-matched product amplitude
\[
\Phi(x,y) = \sqrt{\rho_x(x)\rho_y(y)}\, e^{i\theta(x,y)}, \tag{5.8}
\]
which carries $\psi$'s own phase field but has the decorrelated amplitude. With this choice, the phase contribution to the kinetic energy is identical for $\psi$ and $\Phi$ and cancels exactly, leaving
\[
\langle T \rangle_\psi - \langle T \rangle_\Phi = \frac{\hbar^2}{8m} J_{\mathrm{corr}}[\rho] \ge 0, \tag{5.9}
\]
with nothing left over from the phase. In fact the amplitude and phase pieces of $\langle T\rangle$ split apart cleanly for \emph{any} $\psi=\sqrt\rho\,e^{i\theta}$:
\[
\langle T \rangle_\psi = \frac{\hbar^2}{8m} J[\rho] + \frac{\hbar^2}{2m} \int\!\!\int \rho\, |\nabla \theta|^2\,dx\,dy, \tag{5.10}
\]
and the same split applied to $\Phi$ only removes the amplitude term. Equation (5.9) is therefore the exact complex-wavefunction counterpart of Eq.~(5.7): decorrelating the amplitude alone costs exactly $\hbar^2/8m\cdot J_{\mathrm{corr}}[\rho]$ in kinetic energy, and the phase structure never enters the comparison. One further caution: if instead we compare $\psi$ against the \emph{fully} phaseless reference $\Phi_0=\sqrt{\rho_x\rho_y}$, the difference $\langle T\rangle_\psi-\langle T\rangle_{\Phi_0}$ picks up an extra, non-negative term $\frac{\hbar^2}{2m}\int\!\!\int\rho|\nabla\theta|^2$, the full phase kinetic energy. That comparison mixes correlation removal with phase removal, and should not be read as a pure decorrelation statement.

Two remarks point beyond what this paper takes on. First, a more parsimonious argument gives a stationary, single-mode version of Eq.~(5.3) directly: minimize $\langle T\rangle$ by itself, subject only to normalization, and ordinary calculus of variations returns $-\Delta\psi \propto \psi$ -- the time-independent eigenvalue equation, with no time dependence anywhere in it. That is a different, smaller claim than a law of motion, not a shortcut to one, and Part II is explicit about the distinction. Second, everywhere above $\psi$ was diagnostic: given a wavefunction, $J[\rho]$ was computed from it. Part II shows the same functional run the other way -- extremizing an action built from $J[\rho]$ generates the full time-dependent Schr\"odinger equation itself, spin included, rather than merely describing a state that already solves it.

\section{An exactly solvable example}
\label{sec:example}

We now work through a fully closed-form example.

Consider the ground state of two coupled quantum harmonic oscillators (say with a bilinear coupling term $-\lambda xy$ in the Hamiltonian) \cite{braunstein}. Its Born density is a bivariate Gaussian with unit variances and correlation coefficient $R$,
\[
\rho(x,y) = \frac{1}{2\pi\sqrt{1-R^2}} \exp\left( -\frac{x^2 - 2Rxy + y^2}{2(1-R^2)} \right), \qquad |R| < 1.
\]
The marginals are standard normal, $\rho_x(x) = \rho_y(y) = (2\pi)^{-1/2} e^{-x^2/2}$, and don't depend on $R$ at all -- a standard property of the bivariate Gaussian \cite{coverthomas}. The covariance matrix is $\Sigma=\begin{pmatrix}1&R\\R&1\end{pmatrix}$, with $\det\Sigma=1-R^2$.

\paragraph{Entropy correlation.} The differential entropy of a bivariate Gaussian is $h[\rho] = \tfrac{1}{2}\log[(2\pi e)^2 \det \Sigma] = \tfrac{1}{2}\log[(2\pi e)^2 (1-R^2)]$, while $h[\rho_x] + h[\rho_y] = \log(2\pi e)$. Subtract, and
\[
\scriptI[\rho] = -\frac{1}{2}\log(1-R^2).
\]
Notice it only depends on $R$, it's zero at $R=0$, and it blows up (logarithmically) as $|R|\to1$.

\paragraph{Fisher correlation.} The Fisher information matrix of a Gaussian is $\Sigma^{-1}$, so $J[\rho] = \mathrm{tr}\,\Sigma^{-1} = 2/(1-R^2)$, while each marginal contributes $J[\rho_x] = J[\rho_y] = 1$. So
\[
J_{\mathrm{corr}}[\rho] = \frac{2}{1-R^2} - 2 = \frac{2R^2}{1-R^2}.
\]

\paragraph{Kinetic energy.} By Eq.~(5.9), for the real two-particle wavefunction $\psi=\sqrt\rho$,
\[
\langle T \rangle_\psi - \langle T \rangle_\Phi = \frac{\hbar^2}{4m} \cdot \frac{R^2}{1-R^2}.
\]

\paragraph{Discussion.} Two things jump out of this example. First, for small $R$, $\scriptI[\rho]\approx R^2/2$ while $J_{\mathrm{corr}}[\rho]\approx 2R^2$, so $J_{\mathrm{corr}}\approx 4\scriptI$ to leading order -- a concrete case of the general ``total amount'' versus ``spatial sharpness'' relationship discussed in Section~\ref{sec:discussion}. Second, both quantities diverge as $|R|\to1$, but at different rates -- logarithmically versus as a simple pole -- which tells you the two functionals have genuinely different sensitivities to strong correlation.

\section{Statistical-mechanical interpretation: Jaynes's maximum-entropy principle as minimum correlation}
\label{sec:jaynes}

We now turn to statistical mechanics. We routinely lose detailed information while keeping some larger-scale information fixed. The fundamental transformation $\rho\mapsto\rho_x\rho_y$ is a particularly clean instance of exactly this: it removes dependence while keeping the observed marginals intact. Under decorrelation, using the differential entropy $h$ from Eq.~(2.4),
\[
\Delta h = \scriptI[\rho] \ge 0. \tag{7.1}
\]
The entropy gain equals $\scriptI[\rho]$, the correlation present in $\rho$ that the map $C$ removes. With the marginals held fixed, more entropy means less mutual information. So the maximum-entropy rule can be read as: select the distribution with the least correlation not demanded by the constraints. One caveat: this ``entropy gain'' is relative to the restricted description $\rho(x,y)$ we've chosen -- the underlying fine-grained entropy of a closed system is exactly conserved, while the correlation migrates into degrees of freedom we're not tracking, not that it's destroyed outright. Within that description, though, the excess quantum kinetic energy gives us the local Fisher measure of that same correlation.

Jaynes's own reading of the maximum-entropy rule was simple: pick the least-biased distribution consistent with the physical information you actually have \cite{jaynes1}. Say that information comes as constraints
\[
\int \rho(z) F_a(z)\,dz = f_a, \qquad a = 1,\dots,r, \tag{7.2}
\]
Together with normalization, this gives you the familiar exponential family,
\[
\rho^*(z) = \frac{1}{Z(\lambda)} \exp\left[ -\sum_{a=1}^r \lambda_a F_a(z) \right]. \tag{7.3}
\]
Nothing more has been assumed.

Our decorrelation map gives this rule a new reading. If the marginals $\rho_x$ and $\rho_y$ are held fixed, then
\[
h[\rho_x] + h[\rho_y] - h[\rho] = \scriptI[\rho] \ge 0, \tag{7.4}
\]
and the first two terms are pinned down. So, over distributions $\rho$ with marginals $\rho_x,\rho_y$ held fixed at these same values,
\[
\arg\max_{\rho} h[\rho] = \arg\min_{\rho} \scriptI[\rho] = \rho_x \rho_y. \tag{7.5}
\]
Maximizing entropy and minimizing correlation are, in this setting, literally the same operation.

This suggests the following physical reading of Jaynes's rule:

\begin{center}
\fbox{\begin{minipage}{0.85\textwidth}
\centering
Given physical constraints and a set of possible distributions or related amplitudes satisfying those constraints, select from that set those with the least correlation.
\end{minipage}}
\end{center}

We call this the principle of minimum justified correlation; the word ``justified'' is essential. An interaction, a conservation law, a boundary condition, a symmetry, or fermionic antisymmetry can \emph{require} correlation. When that happens, the extremum distribution will indeed be correlated -- but the correlation entered through the physics of the constraint, not because we snuck in an unjustified extra assumption.

What happens when a single system admits more than one physically relevant state -- as with the excited states of a bound Hamiltonian -- needs more care than the fixed-marginal statement above provides, and is worked out in Part II.

If every constraint splits cleanly into an $x$ part and a $y$ part, the maximum-entropy state factorizes. If a constraint contains a non-separable term $F(x,y)$, the exponential generally will not factorize, and the resulting mutual information is simply the information carried by that coupling constraint.

As already noted in Section~\ref{sec:kinetic}, this whole construction presupposes a physically meaningful split into ``$x$'' and ``$y$'' -- two distinguishable particles, say, or two experimentally distinguished degrees of freedom. For a single particle in more than one dimension, there is no canonically preferred split, and a rotation mixing the coordinate axes will generally change both $\scriptI[\rho]$ and $J_{\mathrm{corr}}[\rho]$, as noted after Eq.~(3.1). The principle of minimum justified correlation is therefore a statement about correlation \emph{relative to a chosen decomposition of the system} -- not an absolute, partition-free statement about a single quantum state.

The uncorrelated state maximizes joint entropy and minimizes correlation-dependent kinetic energy. If constraints require correlation, mutual information tells you how much of it there is, and relative Fisher information tells you how sharply it varies in space. That fixed-marginal result is exact. For more general constraints, the broader statement is a physical interpretation: don't put correlation into the distribution unless the available information actually requires it.

\section{Discussion}
\label{sec:discussion}

We now step back to see why the entropy and kinetic-energy results are, at their core, the same statement told twice. Write the departure from independence as
\[
\eta(x,y) = \log \frac{\rho(x,y)}{\rho_x(x)\rho_y(y)}. \tag{8.1}
\]
Then
\[
\scriptI[\rho] = \int\!\!\int \rho\, \eta\,dx\,dy, \qquad J_{\mathrm{corr}}[\rho] = \int\!\!\int \rho\, |\nabla \eta|^2\,dx\,dy. \tag{8.2}
\]
These are two complementary averages of the very same field $\eta$. $\scriptI[\rho]$ is the mean value of $\eta$ -- a total, dimensionless dependence measure. $J_{\mathrm{corr}}[\rho]$ is the mean squared gradient of $\eta$ -- it measures how fast that dependence varies from point to point. Multiply by $\hbar^2/8m$ and $J_{\mathrm{corr}}$ picks up both the right dimensions and, for a real wavefunction, the exact value of the kinetic-energy difference.

The role of the fundamental map should now be clear. The map $\rho\mapsto\rho_x\rho_y$ holds the marginals fixed and sets $\eta$ to zero. That single move removes \emph{both} the total correlation measured by $\scriptI[\rho]$ and the spatial structure measured by $J_{\mathrm{corr}}[\rho]$, all at once. Entropy goes up by the first amount; Fisher information and kinetic energy go down by the second. These are not three separate analogies; \emph{they are three consequences of one controlled act of removing dependence.}

A word on scope is warranted here. We are \emph{not} deriving quantum mechanics from Shannon entropy, and we are \emph{not} saying all kinetic energy is correlation energy. For a real wavefunction, the kinetic-energy difference follows exactly from the amplitude $\sqrt\rho$ alone. For a complex wavefunction with the phase-matched reference of Eq.~(5.8), the amplitude gives the same non-negative Fisher contribution and the phase drops out cleanly; use the phaseless reference $\Phi_0$ instead, and an extra non-negative phase term shows up, as we saw in Section~\ref{sec:kinetic}. For identical fermions, the simple product amplitude doesn't have to respect antisymmetry, so treat it as a density-level diagnostic, not a competing physical wavefunction. Appendix~\ref{app:lowdin} works through how this compares with the conventional quantum-chemical correlation energy.

In statistical mechanics, the fixed-marginal statement is exact: among all joint distributions with given marginals, the product distribution has maximum entropy and zero mutual information, no hedging required. Interactions, conservation laws, boundary conditions, symmetries, and antisymmetry can all require dependence. The point is never to strip that correlation out -- it's to avoid introducing correlation beyond what those constraints actually demand.

\section{Conclusion}
\label{sec:conclusion}

We conclude by drawing the threads together. We started with the fundamental correlation-destroying map
\[
C : p_{ij} \longmapsto \pi_{ij} = P_i Q_j, \tag{9.1}
\]
which keeps both marginals and removes only the dependence between them, obeying
\[
H(\pi) - H(p) = I(X;Y) \ge 0. \tag{9.2}
\]
The entropy you gain by destroying the correlation is exactly the correlation that was there to begin with. No approximations, no estimation -- $I(X;Y)$ \emph{is} the mutual information.

Move to a continuous density and the same map becomes $\rho\mapsto\rho_x\rho_y$, and the same one map produces the same kind of exact hierarchy in all three currencies at once:
\begin{equation}
\boxed{
\begin{aligned}
h[\rho_x \rho_y] - h[\rho] &= \scriptI[\rho], \\[4pt]
J[\rho] - J[\rho_x] - J[\rho_y] &= J(\rho\|\rho_x\rho_y) = J_{\mathrm{corr}}[\rho], \\[4pt]
\langle T \rangle_\psi - \langle T \rangle_\Phi &= \frac{\hbar^2}{8m} J_{\mathrm{corr}}[\rho] \ge 0 \qquad (\psi = \sqrt{\rho} \text{ real}).
\end{aligned}
}
\tag{9.3--9.5}
\end{equation}
The first line gives the amount of correlation present. The second gives its spatial sharpness. The third is the most striking: that Fisher correlation \emph{is}, exactly, the quantum kinetic energy lost when the dependence is removed from the Born density -- equivalently, exactly the amount by which the von Weizs\"acker functional $T_W[\rho]$ exceeds additivity under marginal factorization. Use a phase-matched product reference for a complex wavefunction, and the amplitude alone delivers this same non-negative Fisher term with no leftover phase contribution, as shown in Eq.~(5.9).

With the marginals fixed, maximizing entropy is identical to minimizing correlation. This gives Jaynes's maximum-entropy principle a complementary physical reading:

\begin{center}
\fbox{\begin{minipage}{0.85\textwidth}
\centering
\textbf{Principle of minimum justified correlation.}\\[2pt]
Given physical constraints and a set of possible distributions or related amplitudes satisfying those constraints, select from that set those with the least correlation.
\end{minipage}}
\end{center}

The word ``justified'' is essential. Interactions, conservation laws, boundary conditions, symmetries, and fermionic antisymmetry can all require correlation, and when they do, that correlation belongs in the physical state -- it is not something to be minimized away. What should not be there is correlation that the constraints do not actually demand. In that sense the map $C$ provides both a precise measure of what dependence contributes, and a concrete standard for judging when it earns its place.

Part II picks up from here: what happens when a system has more than one physically relevant state, what happens when correlation genuinely evolves in time, a generative route to the Schr\"odinger equation itself built from the same Fisher functional, the role of spin, and a proposed experimental test of the dynamical identity.

\subsection*{Acknowledgements}

This research began with the author's PhD study in chemistry at Caltech (1974--1977), supported by an NSF Fellowship; that thesis remains unpublished. He acknowledges the support of his thesis advisor, W. A. Goddard III, and conversations with many people there. The author also acknowledges stimulating conversations with E. T. Jaynes in his office at Washington University in St. Louis. More recent encouragement from K. A. Dill at Stony Brook University has been crucial in stimulating the author to return to this research fifty years later, in sharpening the arguments, and in writing it up for publication. This manuscript was prepared with the assistance of AI, used for literature and citation searching, calculation verification, and iterative drafting and revision of the text under the author's direction.

\appendix

\section{Fisher decomposition}
\label{app:fisher-decomp}

This is the proof promised in Section~\ref{sec:fisher}. Let $s_x = \partial_x \log \rho$ and $\bar{s}_x = \partial_x \log \rho_x$.\\
Since $E[s_x \mid X=x] = \bar{s}_x$,
\[
\int\!\!\int \rho\, |s_x - \bar{s}_x|^2 = \int\!\!\int \rho\, s_x^2 - \int \rho_x\, \bar{s}_x^2. \tag{A.1}
\]
Adding the analogous result for $y$ gives (4.4).

\section{Relation to L\"owdin's correlation energy}
\label{app:lowdin}

Some care is needed here, since ``correlation energy'' already has a specific meaning in quantum chemistry that differs from ours. L\"owdin defined it as the gap between the exact nonrelativistic energy and the Hartree--Fock energy,
\[
E_c^L = E_{\mathrm{exact}} - E_{\mathrm{HF}}. \tag{B.1}
\]
For a ground state, $E_{\mathrm{exact}}\le E_{\mathrm{HF}}$, so this quantity is negative -- though some authors flip the sign and quote a positive ``amount recovered.''

There is an obvious family resemblance between the two constructions: in both cases the physical state is compared against a reference with some correlation removed. But the references, and what is actually being compared, are quite different. The Hartree--Fock reference is the single best Slater determinant. It is \emph{not} uncorrelated in the plain sense $\rho=\rho_x\rho_y$, because antisymmetry has already built exchange into it. In the broader information-theoretic sense used throughout this paper, exchange itself is a form of correlation, so restricting ``correlation energy'' to the gap beyond Hartree--Fock understates the idea; it is really all correlation.

L\"owdin's quantity captures everything left over beyond Hartree--Fock, including changes in both kinetic and interaction energy,
\[
E_c^L = (T_{\mathrm{exact}} - T_{\mathrm{HF}}) + (V_{\mathrm{exact}} - V_{\mathrm{HF}}). \tag{B.2}
\]

Our reference, by contrast, is $\Phi=\sqrt{\rho_x\rho_y}$, built directly from the marginals of the Born density. It keeps those marginals and strips their dependence. We compare kinetic energies only:
\[
T_{\mathrm{corr}}^F = \langle T \rangle_\psi - \langle T \rangle_\Phi = \frac{\hbar^2}{8m} J(\rho\|\rho_x\rho_y) \ge 0. \tag{B.3}
\]
Call this a kinetic Fisher correlation. It has no potential-energy piece in it, and it is positive by construction. It should not be confused with L\"owdin's quantity, which is normally negative for a ground state; there is no contradiction here. Correlation can raise the kinetic energy while lowering the interaction energy by a larger amount, so the total still falls.

\begin{center}
\small
\begin{tabularx}{\textwidth}{@{}l>{\raggedright\arraybackslash}X>{\raggedright\arraybackslash}X@{}}
\hline
 & L\"owdin correlation energy & Fisher kinetic correlation \\
\hline
Reference state & Best Hartree--Fock Slater determinant & Product of the exact density marginals \\
Quantity compared & Total energy, $T+V$ & Kinetic energy only \\
Ground-state sign & $E_c^L \le 0$ & $T_{\mathrm{corr}}^F \ge 0$ \\
Exchange & Already present in Hartree--Fock & Removed by a simple product-density reference \\
Correlation measure & Energy beyond a single determinant & Relative Fisher information of the joint density \\
\hline
\end{tabularx}
\end{center}

Our Fisher-kinetic quantity isn't cleanly separating exchange from other correlation the way L\"owdin's framework does; it's an all-in-one measure that removes exchange effects indiscriminately along with ordinary statistical correlation.

If $x$ and $y$ are electron coordinates, one caution is essential: the simple product amplitude doesn't have to preserve fermion antisymmetry, and it's not a competing electronic-structure wavefunction. Equation (B.3) is a diagnostic at the density level -- it tells you how much kinetic Dirichlet structure is carried by dependence between the chosen variables. A natural next step, not taken here, would be to use an antisymmetric reference -- an optimized determinant, say, or an exchange-preserving pair density -- and ask how much of L\"owdin's correlation energy can be written as a relative Fisher quantity.

The two ideas are complementary. L\"owdin measures the total energetic importance of correlation in a many-electron system. We measure the kinetic structure carried by dependence. What remains open is exactly how the usual correlation energy divides among kinetic, interaction, exchange, and information-geometric parts.

\section{Comparison with Frieden}
\label{app:frieden}

It is worth comparing this construction with that of B. Roy Frieden, who has developed a broad program using Fisher information as a variational basis for physical law -- he calls it extreme physical information, or EPI. In its basic form, an observed effect is described by amplitudes $q_n(x)$, and the data Fisher information is
\[
I = 4\sum_n \int |\nabla q_n(x)|^2\,dx. \tag{C.1}
\]
The information intrinsic to the source is denoted $J$. Frieden assumes that measurement, or some other irreversible transfer, gives $I\le J$, and then derives physical laws by extremizing the loss,
\[
\delta(I - J) = 0, \tag{C.2}
\]
along with additional assumptions about the source and the efficiency of information transfer. Frieden thus uses Fisher information to \emph{generate} physical laws, rather than compute it once the laws are already known.

The point of contact is as follows. If $\rho=q^2$, then
\[
J[\rho] = \int \frac{|\nabla \rho|^2}{\rho}\,dx = 4\int |\nabla q|^2\,dx. \tag{C.3}
\]
This is the same Fisher--Dirichlet identity sitting in Frieden's $I$ and in Eq.~(5.5). Both constructions identify the Dirichlet energy of a probability amplitude with Fisher information, and connect it to the differential structure of quantum mechanics.

But the same expression plays a different role in each argument:

\begin{enumerate}
\item \textbf{Formal correspondence versus physical interpretation.} Frieden ties Fisher information to physical law through a shared mathematical and variational form. Here, the Fisher--Dirichlet expression is given a specific physical meaning: it measures the spatial sharpness of correlation between two variables. Removing that correlation removes exactly this excess, which is why it is non-negative.
\item \textbf{Total information versus correlation information.} Frieden's $I$ is Fisher information in observed data relative to a parameter or source. Ours is the relative Fisher information $J(\rho\|\rho_x\rho_y)$ -- specifically the part due to correlation between the two chosen variables.
\item \textbf{Variational postulate versus exact comparison.} EPI extremizes $I-J$. Equation (5.7) starts from the ordinary quantum kinetic-energy operator and compares two explicitly constructed states -- no new extremum postulate needed.
\item \textbf{Relation to Shannon entropy.} Frieden leans on Fisher information rather than Shannon entropy. We keep both: mutual information for total correlation, relative Fisher information for its local sharpness.
\item \textbf{Scope of the claim.} EPI aims at broad classes of physical law. Ours is narrower: for a given wavefunction and a chosen split of variables, the kinetic energy removed by decorrelation is a Fisher measure of that correlation.
\end{enumerate}

The two programs are therefore complementary rather than competing. Frieden supplies the broad precedent for treating a Fisher--Dirichlet expression as physical information. We isolate one specific part of it and identify it with correlation, tying it back to Shannon mutual information. In EPI's own language, we have picked out a correlation channel within the acquired information. None of this leans on Frieden's source functional $J$, or on Eq.~(C.2).

\section{Relation to Caticha's entropic dynamics}
\label{app:caticha}

A word is also due on Caticha's entropic dynamics, which builds quantum mechanics up from maximum-entropy updating subject to physical constraints. In spirit, this is close to the Jaynesian reading of Section~\ref{sec:jaynes}. Caticha starts from a prior with no correlations at all; directional motion and correlations, including entanglement, are introduced through further constraints as the construction proceeds.

The present construction is narrower and more modest. We start with a state that is already given, remove its correlation by marginal factorization, and measure exactly what came out. The Shannon and Fisher differences are non-negative, and the Fisher difference shows up directly in the quantum kinetic energy. We are measuring the correlation a state already carries, not attempting to rebuild quantum dynamics from scratch.


\begin{thebibliography}{99}

\bibitem{vandrie2000a} J. H. Van Drie, ``The Boltzmann/Shannon entropy as a measure of correlation,'' arXiv:math-ph/0001024 (2000).

\bibitem{vandrie2000b} J. H. Van Drie, ``A Generalization of Jaynes' Principle: An Information-Theoretic Interpretation of the Minimum Principles of Quantum Mechanics and Gravitation,'' arXiv:math-ph/0001028 (2000).

\bibitem{feynman1949} R. P. Feynman, ``The Theory of Positrons,'' Physical Review \textbf{76}, 749--759 (1949).
\bibitem{jaynes1} E. T. Jaynes, ``Information Theory and Statistical Mechanics,'' Physical Review \textbf{106}, 620--630 (1957).

\bibitem{jaynes2} E. T. Jaynes, ``Information Theory and Statistical Mechanics. II,'' Physical Review \textbf{108}, 171--190 (1957).

\bibitem{shannon} C. E. Shannon, Bell System Technical Journal \textbf{27}, 379--423, 623--656 (1948).

\bibitem{coverthomas} T. M. Cover and J. A. Thomas, \emph{Elements of Information Theory}, 2nd ed. (Wiley, 2006).

\bibitem{sears} S. B. Sears, R. G. Parr, and U. Dinur, Israel Journal of Chemistry \textbf{19}, 165--173 (1980).

\bibitem{hamilton} I. P. Hamilton and R. A. Mosna, J. Comput. Appl. Math. \textbf{233}, 1542--1547 (2010).

\bibitem{fisher1925} R. A. Fisher, ``Theory of Statistical Estimation,'' Proceedings of the Cambridge Philosophical Society \textbf{22}, 700--725 (1925).

\bibitem{braunstein} S. L. Braunstein and P. van Loock, ``Quantum information with continuous variables,'' Reviews of Modern Physics \textbf{77}, 513--577 (2005).

\bibitem{frieden1} B. R. Frieden, \emph{Physics from Fisher Information: A Unification} (Cambridge University Press, 1998).

\bibitem{frieden2} B. R. Frieden and B. H. Soffer, ``Lagrangians of physics and the game of Fisher-information transfer,'' Physical Review E \textbf{52}, 2274--2286 (1995).

\bibitem{lowdin} P.-O. L\"owdin, ``Quantum Theory of Many-Particle Systems. III,'' Physical Review \textbf{97}, 1509--1520 (1955).

\bibitem{caticha1} A. Caticha, ``Entropic Dynamics, Time and Quantum Theory,'' Journal of Physics A \textbf{44}, 225303 (2011).

\bibitem{caticha2} A. Caticha, ``Entropic Dynamics: Quantum Mechanics from Entropy and Information Geometry,'' Annalen der Physik \textbf{531}, 1700408 (2019).

\end{thebibliography}
\end{document}